\documentclass[12pt,a4paper]{article}
\usepackage{latexsym}
\usepackage{epsf}
\usepackage{amssymb}

\makeatletter
\@addtoreset{equation}{section}
\makeatother

\begin{document}

\begin{flushright}
\small
IFT-UAM/CSIC-02-40\\
{\bf gr-qc/0210039}\\
October 12th 2002\\
\normalsize
\end{flushright}

\begin{center}


\vspace{.7cm}

{\Large {\bf Supergravity Vacua Today}}\footnote{Talk given at the 
{\it Spanish Relativity Meeting (``EREs'') 2002}, Ma\'o, Menorca, 
September 21-23  2002.}\\

\vspace{1.2cm}

{\bf\large Natxo Alonso-Alberca}${}^{\spadesuit,\heartsuit,}$
\footnote{E-mail: {\tt natxo@leonidas.imaff.csic.es}}
{\large and}
{\bf\large Tom\'as Ort\'{\i}n}${}^{\spadesuit,\clubsuit,}$
\footnote{E-mail:  {\tt Tomas.Ortin@cern.ch} }
\vskip 1truecm

${}^{\spadesuit}$\ {\it Instituto de F\'{\i}sica Te\'orica, C-XVI,
Universidad Aut\'onoma de Madrid \\
E-28049-Madrid, Spain}

\vskip 0.2cm
${}^{\heartsuit}$\ {\it Departamento de F\'{\i}sica Te{\'o}rica, C-XI,
Universidad Aut\'onoma de Madrid\\
Cantoblanco, E-28049 Madrid, Spain}

\vskip 0.2cm
${}^{\clubsuit}$\ {\it I.M.A.F.F., C.S.I.C., Calle de Serrano 113 bis\\ 
E-28006-Madrid, Spain}

\vspace{.7cm}


{\bf Abstract}

\end{center}

\begin{quotation}

\small

We review the definition of (maximally supersymmetric) vacuum in
supergravity theories, the currently known vacua in arbitrary
dimensions and how the associated supersymmetry algebras can be found.

\end{quotation}

\newpage

\pagestyle{plain}

\section*{Introduction}

The vacuum is the most important state of any QFT. It can be defined
as the state with lowest energy and maximal symmetry and it determines
the kinematics of the theory due to the fundamental relation between
symmetries, conserved quantities, quantum numbers and spectrum.

In theories of gravity, the vacuum also determines the zero point of
the energy. Usually, it is associated to a classical solution of
maximal symmetry which can be used as the arena on which other field
theories can be defined. The basic example is Minkowski spacetime.

Some theories (in particular candidates to ``Theory of Everything'')
admit more than one vacuum and the problem of the vacuum selection
(one of whose manifestations is the so-called ``moduli problem'') is
the most pressing and interesting one.

In this talk we are going to focus on vacua of supergravity (SUGRA)
theories. These theories (we do not know if all of them) represent the
low-energy effective limit of a Superstring Theory and, at the same
time, can be considered as simply GR coupled to specific matter fields

\begin{displaymath}
{\rm GR+matter} \longrightarrow \,\,\,{\rm SUGRA}\,\,\,
\stackrel{low\, energy}{\longleftarrow} {\rm Superstrings} 
\end{displaymath}

SUGRA vacua are, to certain approximation, superstring vacua and
provide new interesting GR solutions. In this talk, after a brief
introduction to SUGRA theories (Section~\ref{sec-sugra}) we review the
known vacua of SUGRA theories and explain how to find the associated
supersymmetry algebras (Section~\ref{sec-solutions}). Finally, we
review the very few known general results on supersymmetric solutions
of SUGRA theories (Section~\ref{sec-results}).

\section{SUGRA Theories}
\label{sec-sugra}

\subsection{Supersymmetry}

Supersymmetry (SUSY) is the ultimate symmetry allowed by the S~matrix
\cite{Haag:1974qh}. It interchanges matter (fermions) with radiation
(bosons, the carriers of interactions), implying a higher level of
unification. In QFT, it interchanges bosonic fields $B$ (tensors) with
fermionic (anticommuting) fields $F$ (spinors). This implies that it
is generated by spinorial, anticommuting {\it supercharges}
$Q^{\alpha}$, and the infinitesimal parameters of the field
transformations are anticommuting spinors $\epsilon^{\alpha}$. By
dimensional arguments \cite{VanNieuwenhuizen:ae}, SUSY transformations
are always of the form
\begin{equation}
\left\{
\begin{array}{rcl}
  \delta_{\epsilon} B &\sim& \bar{\epsilon} F\, ,\\
  \delta_{\epsilon} F &\sim& \partial \epsilon + B \epsilon\, .\\
\end{array}
\right.
\end{equation}

The simplest superalgebra includes the Poincar\'e algebra
$\{P_{a},M_{ab}\}$ as bosonic (or {\it even} subalgebra and has the
additional (anti-) commutators
\begin{equation}
\label{eq:originalsusy}
  \left\{ Q^{\alpha},Q^{\beta} \right\}
     = i \left( \gamma^{a} {\cal C}^{-1} \right)^{\alpha\beta} P_{a}\, , 
\hspace{1cm}
  \left[ Q^{\alpha}, M_{ab} \right]
     = {\textstyle\frac{1}{2}} \left( \gamma_{ab} \right)^{\alpha}{}_{\beta}
        Q^{\beta}\, .
\end{equation}

SUSY is, therefore a {\it spacetime} symmetry. In fact, it takes its
simplest form as a symmetry of {\it superspace}.

\subsection{From Supersymmetry to SUGRA}

GR can be viewed\footnote{Strictly speaking it is not a gauge theory
  of the Poincar\'e group, but it can be constructed following the
  same steps, up to certain point.} as a gauge theory of the
Poincar\'e group (see e.g.~\cite{Freund:ws}). The gauge potential
${\cal A}_{\mu}$ has one component for each generator
\begin{equation}
  \left\{ M_{ab}\, , P_{a} \right\} \, \hspace{.2cm}
 {\longrightarrow}\, \put(-25,-10,0){\scriptsize{\rm gauging}}\,
  \hspace{.2cm}
  {\cal A}_{\mu} \equiv {\textstyle\frac{1}{2}} \omega_{\mu}{}^{ab} M_{ab}
  + e_{\mu}{}^{a} P_{a}\, ,
\end{equation}

\noindent 
as well as the curvature
\begin{equation}
 R_{\mu\nu} \equiv 2\, \partial_{[\mu} {\cal A}_{\nu]} 
+\left[ {\cal A}_{\mu}, {\cal A}_{\nu} \right]
 = {\textstyle\frac{1}{2}} R_{\mu\nu}{}^{ab} M_{ab} 
+R_{\mu\nu}{}^{a} P_{a}\, .
\end{equation}

\noindent 
The components $R_{\mu\nu}{}^{ab}$ are the Lorentz curvature and
$R_{\mu\nu}{}^{a}$ to the torsion.  The action is the first-order
Einstein-Hilbert action
\begin{equation}
 S \sim \int d^{4}x\, e\, R(e,\omega)\, ,
\label{eq:action-einstein}
\end{equation}

\noindent 
and the equations of motion that follow from (\ref{eq:action-einstein}) are
fully equivalent to Einstein's vacuum equations
\begin{equation}
  R_{\mu\nu}{}^{a} = 0\, , 
\hspace{1cm}
  G_{\mu\nu} = 0\, ,
\end{equation}

\noindent
although this formulation allows for the coupling of fermions to
gravity (the Cartan-Sciama-Kibble (CSK) theory, see,
e.g.~\cite{DeSabbata:sv,Hehl:kj}).

Similarly, SUGRA can be seen as the gauge theory of the Poincar\'e
{\it superalgebra}. The gauge potential has one more component
$\psi_{\mu}{}^{\alpha}$
\begin{equation}
  \left\{ M_{ab}\, , P_{a}\, , Q^{\alpha}\right\} \, \hspace{.2cm}
 {\longrightarrow}\, \put(-25,-10,0){\scriptsize{\rm gauging}}\,
  \hspace{.2cm}
  {\cal A}_{\mu} \equiv {\textstyle\frac{1}{2}} \omega_{\mu}{}^{ab} M_{ab}
  + e_{\mu}{}^{a} P_{a} + \bar{\psi}_{\mu\alpha} Q^{\alpha}\, .
\end{equation}

\noindent
that compensates for local SUSY transformations and is known as the
Rarita-Schwinger field, that describes a massless spin-$3/2$ particle:
the {\it gravitino}. Its two possible helicity states ($\pm 3/2$) are
the superpartners of the two helicity states of the graviton ($\pm
2$). The QFT will have the same number of bosonic and fermionic states
at each mass level, a property of linearly realized SUSY.

The curvature is
\begin{equation}
 R_{\mu\nu} \equiv 
2\, \partial_{[\mu} {\cal A}_{\nu]} 
+\left[ {\cal A}_{\mu}, {\cal A}_{\nu} \right]
 = {\textstyle\frac{1}{2}} R_{\mu\nu}{}^{ab} M_{ab} 
+R_{\mu\nu}{}^{a} P_{a} + \bar{R}_{\mu\nu\alpha} Q^{\alpha}\, ,
\end{equation}

\noindent
and the action ($N=1,d=4$ SUGRA) is just the CSK theory for a
Rarita-Schwinger field couplet to gravity
\begin{equation}
 S \sim \int d^{4}x\, e\, \left\{ R(e,\omega) +
   \epsilon^{\mu\nu\rho\sigma} \bar{\psi}_{\mu}\gamma_{5}
   \gamma_{\nu} \mathcal{D}_{\rho} (\omega) \psi_{\sigma} \right\}\, ,
\end{equation}

\noindent
but turns out to be invariant under local SUSY transformations
\begin{equation}
\delta_{\epsilon} e^{a}{}_{\mu} =
-i\bar{\epsilon}\gamma^{a}\psi_{\mu}\, ,
\hspace{1cm}
\delta_{\epsilon} \psi_{\mu} =
{\cal D}_{\mu}\epsilon\, .
\end{equation}

Now there is non-vanishing torsion, proportional to the fermions
\begin{equation}
 T_{\mu\nu}{}^{a} \sim \bar{\psi}_{\mu} \gamma^{a} \psi_{\nu}\, .
\end{equation}

Setting all the fermions to zero is always a consistent truncation and
any purely bosonic GR solution will also be a solution of $N=1,d=4$
SUGRA.

Generalizing the superalgebra we gauge we can generalize the SUGRA
theory\footnote{Of course, we can always add supersymmetric matter,
  but here we are not interested in this possibility.}  There are
three main ways (that can be combined) to generalize the Poincar\'e
superalgebra:

\vspace{.3cm}
\noindent 
{\bf 1 Adding more supercharges}

\noindent
Adding more supercharges $Q^{i\alpha}$, $i=1\ldots N$, one is left
with $N$-extended $d=4$ Poincar\'e superalgebras. These turn out to
admit {\it central charges}
\begin{equation}
 Q^{ij} = - Q^{ji}\, , \hspace{.3cm} P^{ij} = -P^{ji}\, ,
\end{equation}

\noindent
that appear in the anticommutator of two supercharges
\begin{equation}
\left\{ Q^{i\alpha}\, , Q^{j\beta} \right\}\,
      = i \delta^{ij} (\gamma^{a} {\cal C}^{-1})^{\alpha\beta} P_{a}
      - i({\cal C}^{-1})^{\alpha\beta} Q^{ij}
      - (\gamma_{5}{\cal C}^{-1})^{\alpha\beta} P^{ij}\, ,
\end{equation}

\noindent
and commute with all generators.  Gauging them we obtain $N$-extended
Poincar\'e SUGRAS. The gauge potential contains $N$ gravitini
$\psi_{\mu}{}^{i\, \alpha}$ and also $N(N-1)/2$ Abelian vector fields
$A^{ij}{}_{\mu}$
\begin{equation}
  {\cal A}_{\mu} \equiv {\textstyle\frac{1}{2}} \omega_{\mu}{}^{ab} M_{ab}
  + e_{\mu}{}^{a} P_{a}
  + {\textstyle\frac{1}{2}} A^{ij}{}_{\mu} Q^{ij}
  + \bar{\psi}^{i}_{\mu\alpha} Q^{i\alpha}\, .
\end{equation}

\noindent 
The action now contains the kinetic terms of $N$ gravitini and of
$N(N-1)/2$ Abelian vector fields $A^{ij}{}_{\mu}$ with field strengths
$ F^{ij}{}_{\mu\nu} = 2\, \partial_{[\mu} A^{ij}{}_{\nu]}$, but this
is not the whole story, as the counting of bosonic and fermionic
states immediately shows: there are additional scalar fields and
fermionic fields in the theory that cannot be accounted for with our
heuristic formulation\footnote{A more rigorous formulation can be
  found in \cite{Castellani:et}.}. The scalars appear always in a
non-linear $\sigma$-model, couple in a non-trivial fashion to the
vector fields and have no potential.

$N=8$ (in $d=4$) is the maximum \cite{Nahm:1978tg} if we do not want
to deal with the problem of higher spin fields in interaction or more
than one graviton.
 
\vspace{.3cm}
\noindent
{\bf 2 Using a different spacetime bosonic superalgebra}

The natural candidates are those superalgebras with a meaningful
bosonic subalgebra\footnote{These can be understood as deformations of
  the Poincar\'e algebra in which the generators of translations
  $P_{a}$ do not commute. There are other possible deformations, like
  the Heisenberg algebras, that but, so far, no SUGRA has been
  constructed gauging them. Further, we could think of taking the
  product of some spacetime group and other compact group that would
  play the role of internal symmetries, but the corresponding SUGRAs
  arise naturally from the $N$-extended $AdS$ SUGRAs we are going to
  consider.}: $dS$ or $AdS$. $dS$ leads to inconsistent field theories
and thus, we are left with $N$-extended $d=4,AdS$ superalgebras. The
supercharges $Q^{i\alpha}$ transform as spinors under the bosonic
($AdS$) subalgebra generated by the $\hat{M}_{\hat{a}\hat{b}}$'s. On
top of these, we are forced to introduce bosonic $SO(N)$ generators
$T^{ij}=-T^{ij}$ that rotate the supercharges and also appear in their
anticommutator:
\begin{equation}
\begin{array}{rcl}
\left\{ Q^{i\alpha}\, , Q^{j\beta} \right\}\,
    & = & 
i \delta^{ij} {m}^{\hat{a}\hat{b}\, \alpha\beta}\, \hat{M}_{\hat{a}\hat{b}}
        - i({\cal C}^{-1})^{\alpha\beta} T^{ij}\, ,\\
& & \\
\left[ T^{ij}\, , T^{kl} \right] & = &  
\delta^{ik} T^{jl} + \delta^{jl} T^{ik}
- \delta^{il} T^{jk} - \delta^{jk} T^{il}\, ,\\
& & \\
\left[ Q^{i\alpha}\, , T^{jk} \right] & = & 
2\, \delta^{i[j} Q^{k]\alpha}\, ,\\
\end{array}
\end{equation}

\noindent 
Accordingly we have to introduce $SO(N)$ an vector field with
$N(N-1)/2$ components $A^{ij}{}_{\mu}$ and $N$ $SO(N)$-charged
gravitini $\psi_{\mu}{}^{i\alpha}$
\begin{equation}
  {\cal A}_{\mu} \equiv 
{\textstyle\frac{1}{2}} 
\hat{\omega}_{\mu}{}^{\hat{a}\hat{b}} \hat{M}_{\hat{a}\hat{b}}
  + {\textstyle\frac{1}{2}} A^{ij}{}_{\mu} T^{ij}
  + \bar{\psi}^{i}_{\mu\alpha} Q^{i\alpha}\, .
\end{equation}

\noindent
The resulting theory is known as a ``gauged SUGRA'' and for $N=1,2$
the Lagrangian has a negative cosmological constant whose value is
related to the gauge coupling constant. For $N>2$ there also scalars
present and there is a potential which has an extremum at a negative
value and acts as a negative cosmological constant. 

\vspace{.3cm}
\noindent
{\bf 3 Using $d>4$ spacetime bosonic superalgebras}

In the Poincar\'e case this is straightforward but only up to $d=11$,
which is allowed only for $N=1$.  Beyond $d=11$ we run into the same
problems we found in going beyond $N=8$ in $d=4$ \cite{Nahm:1978tg}.
These limits are related since $N=8,d=4$ SUGRAs can be derived from
$N=1,d=11$ SUGRA \cite{Cremmer:1978km} by dimensional
reduction\footnote{A very complete guide to the literature on SUGRAs
  in diverse dimensions is \cite{Salam:fm}.}. $AdS$ superalgebras only
exist up to $d=7$ \cite{Nahm:1978tg}.

The main feature of higher-dimensional superalgebras is that they
admit {\sl quasi-central charges} ${\cal Z}_{[a_{1}\ldots a_{p}]}$
that commute with the supercharges but transform as $p$-forms under
Lorentz transformations.  An important example is the $N=1,d=11$
Poincar\'e superalgebra (a.k.a.~``M-superalgebra'', see
\cite{Townsend:1997wg} for a discussion of the great amount of
information that it contains), which admits two quasi-central charges
$p=2,5$
\begin{equation}
\left\{ Q^{\alpha}\, , Q^{\beta} \right\}\,
= i (\gamma^{a} {\cal C}^{-1})^{\alpha\beta} P_{a}
+{\textstyle\frac{1}{2}} 
(\gamma^{ab} {\cal C}^{-1})^{\alpha\beta} {\cal Z}_{ab}
+{\textstyle\frac{i}{5!}}
 (\gamma^{a_{1}\ldots a_{5}}{\cal C}^{-1})^{\alpha\beta}
{\cal Z}_{a_{1}\ldots a_{5}}\, .
\end{equation}

The gauge potential must include now a potential $C_{\mu}{}^{ab}$
\begin{equation}
  {\cal A}_{\mu} \equiv {\textstyle\frac{1}{2}} 
\omega_{\mu}{}^{ab} M_{ab}
  + e_{\mu}{}^{a} P_{a}
  +{\textstyle\frac{1}{2}} C_{\mu}{}^{ab} {\cal Z}_{ab}
  + \bar{\psi}_{\mu\alpha} Q^{\alpha}\, ,
\end{equation}

\noindent 
which actually appears in the SUGRA action as a 3-form potential
$C_{\mu\nu\rho}$ with field strength $ G_{\mu\nu\rho\sigma} =
4\partial_{[\mu} C_{\nu\rho\sigma]}$.

$(p+1)$-form potentials naturally couple to the worldvolume of
extended objects of $p$ spatial dimensions (``$p$-branes'').
Higher-dimensional SUGRAs, are, therefore, theories associated to
$p$-branes and, indeed, one finds classical solutions that include the
$(p+1)$-dimensional Poincar\'e group in their isometry group and
represent the long-range fields sourced by a flat
$p$-brane\footnote{for a review on $p$-branes see, for example
  \cite{Stelle:nv}.}. $N=1,d=11$ SUGRA can couple to a 2-brane and,
through the dual 6-form potential, to a 5-brane. The quasi-central
charges that appear in the superalgebra correspond to these objects
and there are classical solutions associated to them. One of the main
properties of these solutions is that they are supersymmetric.

\section{From SUGRA Back to SUSY}
\label{sec-solutions}

In the previous section we have seen heuristically how to construct
SUGRA theories based on a superalgebra. Most SUGRAs (actions and
transformation laws), though, have not been constructed in this way
and the question arises as to what superalgebra can be associated to a
given SUGRA. 

Actually, it is the vacuum of the theory what can be connected to a
superalgebra and it so happens that many SUGRAs admit several vacua.
Here we are going to show how to find the superalgebra associated to a
given SUGRA vacuum. We need first a definition of vacuum: in GR, the
vacuum is the maximally symmetric solution (i.e.~Minkowski, $AdS$ or
$dS$). In SUGRA we can define it as a maximally supersymmetric
solution.

\subsection{Supersymmetric Solutions}

GR is invariant under any diffeomorphism but any given solution is
only invariant under a finite {\it isometry group} generated by its
Killing vectors $k_{(I)}$, that satisfy a {\sl Lie algebra}
\begin{equation}
 [k_{(I)}\, , k_{(J)}] = f_{IJ}{}^{K} k_{(K)}\, .
\end{equation}

We can call {\it symmetric} any solution with a non-trivial isometry
group.

A SUGRA theory is invariant under any local supersymmetry
transformation, but a given solution will only be invariant under
some. By analogy, we call a SUGRA solution {\it supersymmetric} if
there exists a (Killing) spinor $\kappa$ such that
\begin{equation}
\delta_{\kappa}=0\, \Longrightarrow\,
\left\{
\begin{array}{rcl}
  \delta_{\kappa} B &=& \bar{\kappa} F = 0\, ,\\
  \delta_{\kappa} F &=& \partial \kappa + B \kappa = 0\, .\\
\end{array}
\right.
\end{equation}

A supersymmetric solution will always be symmetric (see later). Its
symmetries and supersymmetries must form a {\it supergroup}, and, the
infinitesimal generators must form a superalgebra. 

How can this superalgebra be found?  The answer to this question was
given in \cite{Figueroa-O'Farrill:1999va}:

\begin{center}
{\it\large \underline{RECIPE}}
\end{center}

\begin{enumerate}
\item First we associate each Killing spinor (vector) to an odd (even)
  element of the superalgebra
\begin{equation}
\left.
\begin{array}{rcl}
  \kappa_{(A)}{}^{\alpha}\, \longrightarrow\, Q_{(A)}\,\,\,\, ({\rm Odd}) \\
& & \\
  k_{(I)}{}^{\mu}\, \longrightarrow\, P_{(I)}\,\,\,\, ({\rm Even}) \\
\end{array}
\right\}
\begin{array}{rcl}
&{\rm SUPERALGEBRA}&\, \\
&{\rm GENERATORS}&\, \\
\end{array}
\end{equation}

\noindent 
The superalgebra is determined by the three sets of structure constants
$f_{IJ}{}^{K},f_{AB}{}^{I},f_{AI}{}^{B}$


\item The structure constants $f_{IJ}{}^{K}$ of the even subalgebra are those
of the isometry Lie algebra
\begin{equation}
 [k_{(I)}\, , k_{(J)}] = f_{IJ}{}^{K} k_{(K)}\, .
\end{equation}

\item The structure constants $f_{AB}{}^{I}$ are given by the decomposition of
the bilinears
\begin{equation}
 -i \bar{\kappa}_{(A)} \gamma^{a} \kappa_{(B)} e_{a} \equiv 
f_{AB}{}^{I} k_{(I)}\, .
\end{equation}

\item The structure constants $f_{AI}{}^{B}$ are given by the {\it
    spinorial Lie derivatives}
\begin{equation}
 \mathbb{L}_{k_{(I)}} \kappa_{(A)} \equiv f_{AI}{}^{B} \kappa_{(B)}\, .
\end{equation}
\end{enumerate}

Some of these rules deserve a comment:

\begin{itemize}
\item {\sl Rule 3}: If $\kappa_{1}$ and $\kappa_{2}$ are Killing
  spinors then one can show that the bilinear
  $-i\bar{\kappa}_{1}\gamma^{\mu}\kappa_{2}$ is a Killing vector, and,
  therefore, a linear combination of the $k_{(I)}{}^{\mu}$'s. This
  ensures that the $f_{AB}{}^{I}$'s are well-defined constants.

  
\item {\sl Rule 4}: The spinorial Lie derivative (see
  \cite{Kosmann,Ortin:2002qb})is a particular case of a {\sl
    generalized G-reductive Lie derivative} (see, e.g.~\cite{Godina}).
  In pedestrian/physicist terms it is just a gauge-covariant Lie
  derivative
\begin{equation}
 \mathbb{L}_{v} = \pounds_{v} + W(v)\, ,
\end{equation}

\noindent 
where $\pounds_{v}$ is the standard Lie derivative and $W(v)$ is a
gauge compensator.  Spinors are defined up to local (gauge) Lorentz
transformations and
\begin{equation}
 \pounds_{v} \psi \equiv v^{\mu} \nabla_{\mu} \psi
  +{\textstyle\frac{1}{4}} \nabla_{[\mu} v_{\nu]} \gamma^{\mu\nu} \psi\, .
\end{equation}

\noindent 
Charged spinors are also defined up to local phases and more terms
have to be added to their Lie derivative. One of its main properties
is that, taken w.r.t.~Killing vectors, it transforms Killing spinors
into Killing spinors. This ensures that the $f_{AI}{}^{B}$'s are also
well-defined constants.

\end{itemize}

\subsection{Vacua and Their Superalgebras}

Let us now review the known maximally supersymmetric solutions of the
different SUGRAs. Applying the above recipe we can derive their
symmetry superalgebras.

\subsubsection{Poincar\'e SUGRAs}

Minkowski spacetime is always a solution, maximally symmetric and
supersymmetric. Its symmetry superalgebra is (not surprisingly) the
Poincar\'e superalgebra:
\begin{enumerate}
\item Minkowski's Killing vectors in Cartesian coordinates are
\begin{equation}
\begin{array}{rcl}
 && k_{(a)} = \partial_{a}\, \longrightarrow\, P_{a}\, ,\\
 && k_{ab} = 2\, x_{[a}\partial_{b]}\, \longrightarrow\, M_{ab}\, .\\
\end{array}
\end{equation}

The Killing spinor equation in Cartesian coordinates with
$e_{\mu}{}^{a}=\delta_{\mu}{}^{a}$ is 
\begin{equation}
 \partial_{a}\kappa=0\, ,
\end{equation}

\noindent 
and has four independent solutions
\begin{equation}
 \kappa_{(\alpha)}{}^{\beta} =  
\delta_{(\alpha)}{}^{\beta}\, \longrightarrow\, Q_{(\alpha)}\, .
\end{equation}

\item $\left\{P_{a}\, , M_{ab}\right\}$ satisfy the Poincar\'e algebra.
\item
The Killing spinor bilinears give, trivially
\begin{equation}
 -i \bar{\kappa}_{(\alpha)} \gamma^{a} \kappa_{(\beta)} \partial_{a}
 = -i ({\cal C}\gamma^{a})_{\alpha\beta} k_{(a)}\,
 \Rightarrow\,
 \left\{Q_{(\alpha)}\, , Q_{(\beta)}\right\} = 
-i ({\cal C}\gamma^{a})_{\alpha\beta} P_{(a)}\, .
\end{equation}

\item 
The translational Killing vectors are covariantly constant and they 
annihilate the Killing spinors
\begin{equation}
 \nabla_{\mu}k_{(a)\nu} = 0\,
 \Rightarrow\, \mathbb{L}_{k_{(a)}}\, \kappa_{(\alpha)}{}^{\beta} = 0\,
 \Rightarrow\, [Q_{(\alpha)}\, , P_{(a)}] = 0\, .
\end{equation}

\noindent 
The rotational Killing vectors have non-trivial covariant derivatives
and act as Lorentz rotations on the Killing spinors
\begin{equation}
  \begin{array}{rcl}
 \nabla_{[\mu}k_{(ab)\nu]} = 2\, e_{[a|\mu} e_{|b]\nu}\,
& \Rightarrow & \mathbb{L}_{k_{(ab)}}\, \kappa_{(\alpha)}{}^{\beta}
 = {\textstyle\frac{1}{2}} 
(\gamma_{ab})^{\beta}{}_{\gamma} \kappa^{\gamma}{}_{(\alpha)}\, ,\\
& & \\
& \Rightarrow &
[Q_{(\alpha)}\, , M_{ab}] 
=Q_{(\beta)} {\textstyle\frac{1}{2}} (\gamma_{ab})^{\beta}{}_{\alpha}\, .\\
\end{array}
\end{equation}

\end{enumerate}

We could raise now the spinorial indices using the inverse charge
conjugation matrix to have the algebra in the form
Eq.~(\ref{eq:originalsusy}).

\subsubsection{Anti-de Sitter Supergravities}

$AdS$ is always a maximally symmetric and supersymmetric solution of
all gauged supergravities. Its symmetry superalgebra can be derived
most easily using a construction that exploits the fact that it is a
homogeneous space that we are going to review in
Section~\ref{sec-GKS}.

\subsubsection{Extended Poincar\'e Supergravities}

Minkowski is always a maximally supersymmetric solution of all
extended Poincar\'e SUGRAS, but there are additional maximally
supersymmetric (but not maximally symmetric) solutions of essentially
two types: products of $AdS$ spaces and spheres that appear as
near-horizon limits and homogenous $pp$-wave spacetimes
\cite{Cahen,Figueroa-O'Farrill:2001nz} of the type found by
Kowalski-Glikman \cite{Kowalski-Glikman:im,Kowalski-Glikman:wv} (KG
solutions) and which are related by a Penrose limit
\cite{kn:Pen6,Gueven:2000ru,Blau:2002dy,Blau:2002rg}.  A complete
table follows:
\begin{displaymath}
\begin{array}{rccc}
N=1\, ,\, d=11:\, &
 \left\{
 \begin{array}{c}
   AdS_{7}\times S^{4}\\
   AdS_{4}\times S^{7}\\
 \end{array}
 \hspace{.6cm}
 \right\}\,
 \put(-30,2,0){\cite{Freund:1980xh}}
& 
\put(-9,3){\vector(1,0){50}}
\put(-9,-10,0){\scriptsize{\rm Penrose\, limit}}\,
&

\hspace{1.3cm}\, KG11\, 
\cite{Kowalski-Glikman:wv}\, \\
 & & \\
N=2B\, ,\, d=10:\, &
 \left\{
   AdS_{5}\times S^{5}
 \right\}\,
& 
\put(-9,3){\vector(1,0){50}}
\put(-9,-10,0){\scriptsize{\rm Penrose\, limit}}\,
&
\hspace{1.3cm}
KG10\,
\cite{Blau:2001ne}\, \\
 & & \\
N=\left(
  \begin{array}{c}
    2,0\\
    4,0\\
  \end{array}
 \right),\,
d=6:\, &
 \left\{
   AdS_{3}\times S^{3}
 \right\}\,
&
\put(-9,3){\vector(1,0){50}}
\put(-9,-10,0){\scriptsize{\rm Penrose\, limit}}\,
&
\hspace{1.3cm}
KG6\,
\put(-10,-45){\vector(0,-1){20}}
\put(-18,-38,0){\cite{Meessen:2001vx}}
\put(-10,-25){\vector(0,1){20}}
\vspace{-1.1cm}
\, \\
 & & \\
N=2\, ,\, d=5:\, &
 \left\{
 \begin{array}{c}
   AdS_{3}\times S^{2}\\
   AdS_{2}\times S^{3}\\
   AdS_{2}*S^{2}\\
   \mbox{G\"odel}\\
 \end{array}
 \hspace{.6cm}
 \right\}
\put(-32,15,0){\cite{Chamseddine:1996pi}}
& 
\put(-9,3){\vector(1,0){50}}
\put(-9,-10,0){\scriptsize{\rm Penrose\, limit}}\,
\put(35,-19){\vector(-1,0){90}}\,
\put(40,-20,0){\rm new}\,
&
\hspace{1.3cm}
KG5\,
\cite{Gauntlett:2002nw}\, \\
 & & \\
N=2\, ,\, d=4:\, &
 \left\{
   AdS_{2}\times S^{2}\, \cite{Bertotti,Robinson}
 \right\}\,
& 
\put(-9,3){\vector(1,0){50}}
\put(-9,-10,0){\scriptsize{\rm Penrose\, limit}}\,
&
\hspace{1.3cm}
KG4\,
\cite{Kowalski-Glikman:im}\, \\
\end{array}
\end{displaymath}

The only two exceptional cases occur in $d=5$: the $AdS_{2}*S^{2}$
solution which is the near-horizon limit of the rotating extreme $d=5$
black hole
\cite{Cvetic:1998xh,Gauntlett:1998fz,Lozano-Tellechea:2002pn} and a
G\"odel-like solution \cite{Gauntlett:2002nw}.

\subsection{The Symmetry Superalgebras of Homogeneous Spacetimes}
\label{sec-GKS}

All known maximally supersymmetric SUGRA vacua are homogeneous spaces.
The less well known cases are
\begin{equation}
\begin{array}{rcl}
AdS_{2} * S^{2} & \sim & \frac{\left[ SO(2,1)\times SO(3) \right]}{SO(2)_{\alpha}}\,
   \hspace{.2cm} \cite{Alonso-Alberca:2002wr},\\
 & & \\
Hpp & \sim & \frac{H(d-2)}{T_{d-2}}\, \hspace{.2cm} \cite{Cahen}.\\
 & & \\
\mbox{G\"odel}_{5} & \sim & \frac{H(2n+2)}{U(1)} \\
\end{array}
\end{equation}

\noindent
where $H(2n+2)$ stands for the Heisenberg algebra.  This makes easy to
find the Killing spinors and supersymmetry algebras.  It can be shown
\cite{Alonso-Alberca:2002gh} that
\begin{enumerate}
\item The Killing spinors are (in an appropriate basis!)
 \begin{equation}
  \kappa_{(\alpha)}{}^{\beta} = u^{\beta}{}_{\alpha}\, ,
\hspace{1cm}
 u = e^{x^{a} \Gamma_{s}(P_{(a)})}\, ,
\end{equation}
\noindent
where $\Gamma_{s}(P_{(a)})$ stands for the generators in the
spinorial representation.

\item In most cases, the bilinears can be decomposed easily
 \begin{equation}
  -i \bar{\kappa}_{(\alpha)} \gamma^{a} \kappa_{(\beta)} e_{a}
  = -i (\tilde{\cal C}\Gamma_{s}(T^{(I)}))_{\alpha\beta} k_{(I)}\, ,
\,\,\, \Rightarrow f_{\alpha\beta}{}^{I} =  
-i (\tilde{\cal C}\Gamma_{s}(T^{(I)}))_{\alpha\beta}\, .
 \end{equation}
 
\item The spinorial Lie derivatives w.r.t.~the Killing vectors are
  given by
 \begin{equation}
  \mathbb{L}_{k_{(I)}} \kappa_{(\alpha)}{}^{\beta}
  = \kappa_{(\gamma)}{}^{\beta} \Gamma_{s} (T_{(I)})^{\gamma}{}_{\alpha}\, ,
\,\,\, \Rightarrow f_{\alpha I}{}^{\gamma} 
=  \Gamma_{s} (T_{(I)})^{\gamma}{}_{\alpha}\, .
 \end{equation}

\end{enumerate}

\section{Some General Results on Supersymmetric Solutions}
\label{sec-results}

To end, let us review the few general results that exist on
(not necessarily maximally) supersymmetric solutions.

\subsection{$N=1$, $d=4$ Poincar\'e Supergravity}

The Killing spinor equation is
\begin{equation}
 \nabla_{\mu} \kappa = 0\,
 \Rightarrow\,
 -{\textstyle\frac{1}{4}} R_{\mu\nu}{}^{ab} \gamma_{ab} \kappa =0\, .
\end{equation}

Only two kinds of solutions known: Minkowski spacetime (maximally
supersymmetric) and $pp$-waves spacetimes admitting a covariantly
constant null Killing vector $\ell^{\mu}$ , whose Killing spinors
satisfy
\begin{equation}
  \ell_{\mu} \gamma^{\mu} \kappa = 0\, ,
\end{equation}

\noindent 
that has a 2-dimensional space of solutions ($1/2$ out of the 4 possible).

In Euclidean signature also the Gibbons-Hawking metrics
\cite{Gibbons:zt} used to construct self-dual gravitational instantons
preserve $1/2$ of the supersymmetries
 \begin{equation}
 \left\{
 \begin{array}{l}
  ds^{2} = H^{-1} (d\tau + A_{i} dx^{i})^{2} + H dx^{i} dx^{i}\, ,\\
  \epsilon_{ijk} \partial_{j} A_{k} = \partial_{i} H\,
 \Rightarrow\, \partial_{i} \partial_{i} H = 0\, .\\
 \end{array}
 \right.
 \end{equation}

\subsection{$N=2,d=4$ Poincar\'e Supergravity}

This theory is just Einstein-Maxwell coupled to $\psi_{\mu}{}^{i}$ and
all solutions of Einstein-Maxwell are solutions of $N=2,d=4$
Poincar\'e supergravity as well.  The Killing spinor equation is
\begin{equation}
 \left[ \delta^{ij} \nabla_{\mu} + {\textstyle\frac{1}{4}}\not\! F
 (\sigma^{2})^{ij} \right] \kappa^{j} = 0\,
 \Rightarrow\,
\left\{ C_{\mu\nu}{}^{ab} \gamma_{ab} + 2i \not\!\!\nabla
 (F_{\mu\nu} + i { }^{*} F_{\mu\nu} \gamma_{5}) i \sigma^{2} \right\} \kappa = 0\, .
\end{equation}

These integrability conditions were fully solved by Tod \cite{Tod:pm},
who found two kind of solutions that preserve generically one half of
the supersymmetries:

\begin{itemize}
\item {\bf Israel-Wilson-Perj\'es solutions}
 \begin{equation}
 \left\{
 \begin{array}{l}
  ds^{2} = |{\cal H}|^{-2} (dt + \omega)^{2} 
- |{\cal H}|^{2} dx^{i} dx^{i}\, ,\\
  A_{t} = 2\, {\rm Re}\, {\cal H}\, ,\,\,\,\, 
\tilde{A}_{t} = -2\, {\rm Re}\, {(i\bar{\cal H})}\, ,\\
  \omega = \omega_{i} dx^{i}\, ,\,\,\,
 \epsilon_{ijk} \partial_{j} \omega_{k} = 
\pm\, {\rm Im}\, {(\bar{\cal H} \partial_{k} {\cal H})}\, ,\,\, \Rightarrow 
  \partial_{i} \partial_{i} {\cal H} = 0\, .\\
 \end{array}
 \right.
 \end{equation}
 
 These include the extreme RN ($M^2=Q^2$) black hole (whose
 near-horizon geometry is the maximally supersymmetric
 Bertotti-Robinson spacetime, $AdS_{2}\times S^{2}$), the extreme
 Taub-NUT ($M^2+N^2=Q^2$), KN with $M^2=Q^2$ and multicenter solutions
 \cite{Majumdar,Papapetrou}.
\item {\bf Gravito-electromagnetic $pp$-waves}
 \begin{equation}
 \left\{
 \begin{array}{l}
  ds^{2} = 2 du (dv + Kdu) - 2 d\xi d\bar{\xi}\, ,\\
  F_{\xi u} = \partial_{\xi} C\, ,\, K = {\rm Re} f + \frac{1}{4}|C|^{2}\, ,\\
  \partial_{\bar{\xi}} f = \partial_{\bar{\xi}} C = 0\, .\\
 \end{array}
 \right.
 \end{equation}
 
 The $Hpp$-waves are a particular (maximally supersymmetric) case with
 $K = A_{ij} x^{i} x^{j}$.  $KG4$ is a particular case with
 $A_{ij}\sim\delta_{ij}$.
 
 Kowalski-Glikman proved that Minkowski, Bertotti-Robinson and the
 $KG4$ solutions are the only vacua of $N=2$ supergravity
 \cite{Kowalski-Glikman:im}.

\end{itemize}

\subsection{$N=4$, $d=4$ Poincar\'e Supergravity}

The theory consists of the metric $g_{\mu\nu}$, six vectors
$A_{\mu}{}^{i}$, a complex scalar field $\tau$ (also known as {\sl
  axidilaton}), four gravitini $\psi_{\mu}{}^{i}$ and four dilatini
$\chi^{i}$.

The most general families of supersymmetric solutions were obtained by Tod
\cite{Tod:jf}, who identified several families:

\begin{itemize}
\item {\bf SUPER-Israel-Wilson-Perj\'es solutions}
  \cite{Bergshoeff:1996gg} Include all the IWP spacetimes of $N=2$
  $d=4$, but now none of them seems to be maximally supersymmetric.
\item {\bf Waves ($pp$ and more)} Again, none of them seems to be
  maximally supersymmetric.
\end{itemize}

\subsection{$N=2$, $d=5$ Poincar\'e SUGRA}

This theory is an interesting modification of the 5-dimensional
Einstein-Maxwell theory that leads to very interesting new solutions
like the rotating black string of \cite{Emparan:2001wn}. The action is
\begin{equation}
 S \sim \int d^{4}x\, e\, \left\{ R(e,\omega)
   -{\textstyle\frac{1}{4}} F^2
   -{\textstyle\frac{1}{\sqrt{|g|}}} \epsilon^{\mu\nu\rho\sigma}
     F_{\mu\nu}F_{\rho\sigma}A_{\delta} \right\}\, ,
\end{equation}

\noindent 
where the new Chern-Simons term changes the Maxwell equation.
Recently it was shown in \cite{Gauntlett:2002nw} how to construct all
the supersymmetric solutions of this theory, although not all of them
can be written explicitely.  This method exploits the identities
satisfied by the Killing spinor bilinears
$-i\bar{\kappa}\gamma^{a}\kappa$. The most interesting of the new
results is the presence of a G\"del-like maximally supersymmetric
solution.

these are all the general results known. In higher dimensions and
higher $N$ many supersymmetric solutions are known but no general
classification scheme exists. This seems a very promising direction of
research, worth pursuing.

\section*{Acknowledgements}

T.O.~would like to thank M.M.~Fern\'andez for her continuous support.
This work has been partially supported by the Spanish grant
FPA2000-1584.



\end{document}